\documentclass[11pt]{article}

\usepackage{multirow}
\usepackage{tabularx}
\usepackage{longtable}
\usepackage{graphicx}
\usepackage{multirow}
\usepackage{multicol}
\begin{document}
% this is an alternate method of creating a title
%\hfill\vbox{\hbox{Gius, Mark}
%       \hbox{Cpe 456, Section 01}  
%       \hbox{Lab 1}    
%       \hbox{\today}}\par
%
%\bigskip
%\centerline{\Large\bf Lab 1: Security Audit}\par
%\bigskip
\author{Alan Smith and Manas Gaur\\ Kno.e.sis Center, Wright State University \\ Dayton, Ohio-USA}
\title{What's my age?: Predicting Twitter User's Age using Influential Friend Network and DBpedia}
\maketitle

\section{Abstract}
Social media is a rich source of user behavior and opinions. Twitter senses nearly 500 million tweets per day from 328 million users.An appropriate machine learning pipeline over this information enables up-to-date and cost-effective data collection for a wide variety of domains such as; social science, public health, the wisdom of the crowd, etc. In many of the domains, users demographic information is key to the identification of segments of the populations being studied. For instance, \textit{Which age groups are observed to abuse which drugs?, Which ethnicities are most affected by depression per location?}. Twitter in its current state does not require users to provide any demographic information. We propose to create a machine learning system coupled with the DBpedia graph that predicts the most probable \textit{age} of the Twitter user. In our process to build an age prediction model using social media text and user meta-data, we explore the existing state of the art approaches. Detailing our data collection, feature engineering cycle, model selection and evaluation pipeline, we will exhibit the efficacy of our approach by comparing with the \textit{"predict mean"} age estimator baseline.      

\textbf{Keywords:} Semantics, Linear and Non-Linear Machine Learning, User profiles, Twitter Influencers, Network Analysis, DBpedia, Error-bound Precision. 

\section{Introduction}
Twitter has been an active social event sensing platform for healthcare[16], NFL sports[17], situational awareness[18], consumer behavior[19], political transition[20], etc. Twitter users have been centric to these application scenarios for deriving an intuitive explanation of an event. Understanding the behavior of Twitter users through their likes, comments, and retweets assists data scientists to provide a framework for content personalization. Personalization of content in terms of music, movies, artists, product, news, medicine, therapy, etc. will assist the user to acquire suitable information and build an individualized friend-follower network. Moreover, content personalization for Twitter users based on their social activities has seen increased adhesion in many web-based companies (e.g. Youtube, Amazon, eBay). Personalization  has seen its rise over the past five years, which has directed Twitter users to align their Twitter profile's content with their interest.
In the recent past, \textit{Apple and Amazon} research has seen a lot of interventions in \textit{Human Factors}, which deals with understanding consumer behavior and predict customer retention. Past state of the art approaches to personalization fail to incorporate \textit{age} in assessing consumer/customer/user trends. To better understand user interest, it is essential to segment the population using demographic information. For instance, \textit{Who listens to Ray Charles?}. We seek to take the support of Twitter users' content personalization activity along with their build-up friend network to provide an estimate of their age. Our task of age prediction concerns those users whose profiles lack age description. \\
Our key contributions are as follows: (1) We define a strategic pipeline of data collection, preprocessing, feature engineering, and model selection for developing a system for age estimation. (2.) We provide a one-hop depth-first search strategy for enriching the count of popular Twitter user profiles. (3.) We demonstrate our feature engineering process based on the interest of users and DBpedia categories. We aim to sneak away from traditional survey/interview-based approach for prediction of user demographics. (4.) We experiment our age estimation approach over a sample of 23,120 twitter users with extensive evaluation using cross-validation.  

\section{Related Work}

\subsection{Age estimation as a prediction problem}
Web-based personalization of users based on their buying and browsing patterns have been unraveled with machine learning and human factors research[19]. Surveys and interviews have been a great source of providing personalization by identifying the demographics of the user, such as with \textit{Amazon Mechanical Turk}[21]. These approaches, though productive, lack scalability, considering the number of weblogs generated each day on various web-based platforms. The question on scalability lies in the involvement of humans to provide manual labeling of each user's demographics (age, gender, and location)[4]. Recently, a Twitter platform has diverted the interest of the user based on the Tweets of their followers. One can observe that the user's interest is majorly affected by the Tweets of their followers on Twitter, Facebook, and Youtube [1]. Due to an increased amount of weblogs from the customer base, mean of the aggregated sum of user's followee network [2] have been employed as an approach for age prediction. Moreover, the demographics of the user can be predicted using their web browsing behavior [2] using a probabilistic machine learning. These approaches lack two important facts; first, the user follows that user who share some interest among each other. Interest can be occupation, music, movies, popular person etc. Secondly, studying the browsing pattern of a user alone is vague and random. A recursive and relative study of similarity between the browsing pattern of two users can provide a precise intuition of a user's demographics. Furthermore, human age prediction has been seen from a series of images, where over the year there is a change in the image pixels. In [3], a probabilistic model has been trained for age estimation using the multimedia content. The approach fails to gather practicality in the real world because of 2 reasons; first, it requires a large number of training samples. Second, it does not take into account the current interest of the user. As with the change of age, the interest of the person changes [12]. There has been a consideration of the fact that age prediction is a hard problem[6] from the perspective of text analysis. This is because a male user following female users tends to write tweets using female linguistic styles. In our approach, we see this problem from the perspective of overlapping interest between a user (whose age is to be predicted) and another user (whose age is known). Considering the age estimation problem as a linear regression problem is another approach detailed in [8]. This approach has experimented with blogs/forums, and conversational logs. The supervised feature set generated is influenced by linguistic styles (e.g. POS tagging, Grammars). These features are a shallow representation of user's demographics and tend to generate high mean absolute errors. One of our key contributions is in defining a prudent feature engineering process for efficient learning of our model.   

\subsection{Age estimation as a classification problem}
Age prediction has been considered as a classification problem in order to handle the dynamicity in social media platforms. In [5], the age attribute is considered as a latent attribute of a Twitter user, and models it as an SVM based classification problem. The World Health Organization (WHO) provides a set of age groups predominant in the United States [13]. Range-based classification (range is analogous to age groups) of Twitter user's demographics is not representative of one's age and fails to achieve precision in determining one's age. For instance, under the age group of 17-30, two persons whose ages are 18 and 28 fall into the same cluster. As a result, a personalized recommendation system built over this system will fail to provide apt judgment. An enhancement of the previous approach is to create a finely grained annotation to assign ages to users[6]. This annotated data is used to train a supervised model, classifying users into different age groups. After classification, other features like life stages etc. were used for finer-grained age prediction. 

\subsection {Similar Work}
A patent defining the procedure for the prediction of user's age based on his/her social network has been stated in [7]. It aims to create a social network of users, and given a user in the network whose age is not stated, the system predicts the age utilizing demographics of another user. This approach, though applicable, fails to provide the exact age of the user, because a social network may not be representative of a user's age. For instance, a teenager showing interest in retro music will most probably be surrounded by users greater than teenager's age. Another instance can be political thoughts. It is very less likely that a high school student engaging in politics and business will be linked to users of his/her age. An improvement over this approach is to analyze the diversity in a user's friend network to generate features influencing the user's age.\\
All the past approaches to age estimation dearth in augmenting high provenance general knowledge sources (e.g. Freebase). These knowledge sources possess categories (e.g. genres, location, occupation, hobbies etc.) which can help in establishing the link between different users in social media. Utilizing the content of human-curated knowledge sources and developing a strategic machine learning pipeline can provide the definite age of a user. Precise prediction of age will assist in better understanding user interest and define a wisdom of crowd paradigm for improving individualistic preferences.

\section{Data Collection}

\begin{figure}
\centering
\includegraphics[scale=0.35]{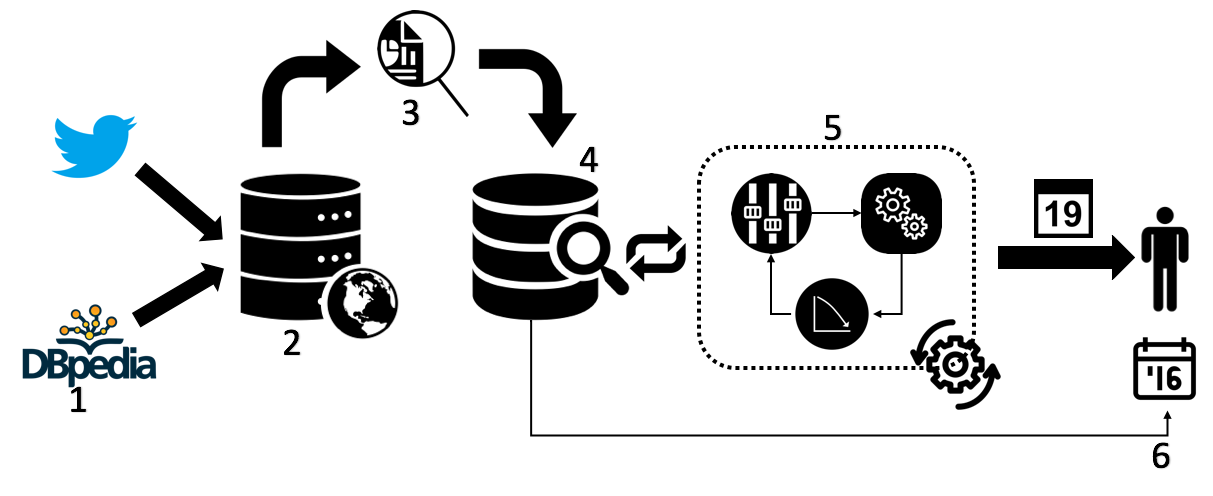}
\caption{Regression based pipeline for Age Prediction. (1). Data Extraction (2). Data Storing (3). Data Analysis (4). Data Pre-Processing (5). Machine learning Pipeline with Hyperparameter Tuning, Model Training and Model Evalution in a Cycle with Different models. (6). Age Prediction by the best model. }
\label{fig:2}
\end{figure}
In our data collection process, we utilized four different data sources: Twitris [24], the Twitter REST API [25], Wikipedia, and DBpedia version 2016-10 [26].
\paragraph{}The Twitris dataset was used as the source of users with self-reported age in their profile description field. This set of users was further filtered by regular expressions, to keep only the users whose age could be extracted with reasonable certainty. We then collected each of these users' friend IDs via the Twitter \textit{friends/ids} API endpoint, where "friend" is defined as a user who is followed by another user. This set of users, with known, extracted age and known friend IDs, represents what was to become the set of observations in our dataset (N=23,120).
\paragraph{}To generate the background knowledge-based features for each observation in the dataset, we first extracted the list of the top 50 most-followed Twitter users from Wikipedia [27], along with the mapping between each user's associated Twitter \textit{screen\_name} and Wikipedia page URL. We then directly translated to the users' Wikipedia URLs the corresponding DBpedia entity URIs. We obtained the "fully-hydrated" user objects for each of the users in the top 50 set using the Twitter API's \textit{users/lookup} endpoint, along with the user objects for each of those popular users' friends, in order to expand our popular users set to include many other \textit{potentially} popular users. On this set of the top 50 users' friends, we used DBpedia Spotlight to link user profiles to their corresponding DBpedia entity URIs, where applicable. To ensure the sufficient accuracy of our entity linking by DBpedia Spotlight, we used a relatively high confidence parameter (0.8), and we annotated the concatenation of the \textit{name} and \textit{description} fields of the user objects for added context to aid in DBpedia Spotlight's disambiguation techniques. However, we considered only annotations found within the \textit{name} field of the user object. In addition to storing the \textit{rdf:type} values for each linked entity, we also queried DBpedia's \textit{Mapping-based Literals} dataset, which included the \textit{DBpedia:birthDate} properties of a significant portion of the entities of type \textit{DBpedia:Person}.
Finally, to form our final feature vectors for each user with known age, we associated each user with his/her set of known popular "friends". We aggregated the counts of each \textit{rdf:type} the user was following (in the DBpedia namespace), along with the mean/median age and \textit{followers\_count} of all the popular users found in the user's friend IDs set. In total, we observed 372 features for each user, including \textit{friends\_count, followers\_count, popular\_friends\_count, mean\_friends\_followers\_count, median\_friends\_followers\_count, mean\_friends\_age, median\_friends\_age,} and the counts of each DBpedia type followed by the user.

\section{Data Analysis}
\begin{figure}[!htbp]
\centering
\includegraphics[scale=0.70]{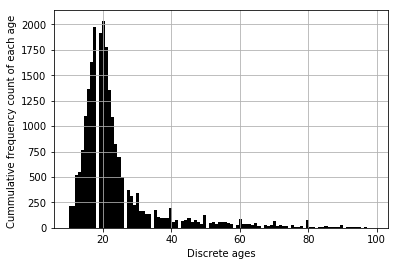}
\caption{Skewness in the data, Bins=100}
\label{fig:1}
\end{figure}

A preliminary analysis using Kolgomorov Smirnov Statistics[11] provided a p-value of 0.04. It suggests that we have 96\% chance of predicting the age of a Twitter user close to their age using the complete set of features. Figure \ref{fig:1}, shows the skewness in the data. Majority of the sample in the dataset represent users in the range [19,29]. A skewed dataset tends to pose serious challenges when modeling the problem as a prediction problem, as opposed to classification problem. In this section, we also provide a descriptive distribution of the age (see table \ref{table:0}).

\begin{table}
\centering
 \begin{tabular}{|c|c|} 
 \hline
 Property & Statistic Value\\ [0.5ex] 
 \hline
 User count & 23120  \\ 
 \hline
 mean. & 23.77  \\
 \hline
 std. & 12.58 \\
 \hline
 min. & 10.00\\
 \hline
 25\% & 17.00 \\
 \hline
 50\% & 50.00 \\
 \hline
 75\% & 25.00 \\
 \hline
 max & 99.00 \\
 \hline
\end{tabular}
\caption{Distribution of Age in the dataset. min. : Minimum age, max.: Maximum age, std.: Standard Deviation, 25\%: First Quartile, 50\%: Median (Second Quartile), 75\%: Third Quartile.}
\label{table:0}
\end{table}

\section{Experimental Staging}
In this section, we will appreciate a set of experiments to validate our claim. Staging of experiments in our approach is necessitated by the high dimension of our features. Recently in [4], machine learning algorithms have suffered from the problem of high variance. We specifically, detailed upon model selection, crowd-inspired classes and evaluation metric selection (see figure \ref{fig:1}).
\subsection{Model Selection}
Selecting an appropriate model is critical for the correct prediction of a user's age. Since we approach the problem by framing it around regression, there is a need for a model that provides precise prediction of age as opposed to classification (the true precision is hidden). For the selection of a suitable model, we utilized cross-validation, Akaike Information Criterion, Bayesian Information Criterion and model-specific hyperparameter tuning. Since our data is skewed (see figure \ref{fig:1}), messy, and imbalanced, there is no straight path for selecting an apt model. Moreover, for some (non-regularized) models, we utilized mutual information for selecting the K best features in our dataset. 
\subsubsection{Akaike Information Criterion (AIC)}
This is a criterion for model selection that is influenced by negative of the log-likelihood of model fitness and model parameters. Though it is a relative measure which compares the fitness of a set of models, we also used it as a guide in tuning the regularization parameter(s) for our models (e.g. LASSO[9]).

\subsection{Cross Validation}
This approach prevents the model from overfitting or underfitting while learning some pattern over the dataset. Since the knowledge-based features in our dataset have smaller $R^2$-score (explained later) compared to count-based, statistical features, we tend to evaluate model fitness using the cross-validation score. Cross-validation requires tuning because of the problem of high variance. For the evaluation of the model, we created a grid of train-test splits for each model, including the test set sizes of \{0.33, 0.25 and 0.1\}.
\subsection{Linear-Models}
Considering the nature of our dataset, we initiated our experiments with linear models.
\paragraph{Linear Regression (LR)} 
This is very naive model employed for age prediction. Linear regression is an algorithm that attempts to find the coefficients in the equation of the best-fit line using the features and observations in the dataset. This model is influenced by the correlation between the target variable and a linear combination of the independent input features. Since, in our dataset, statistical and Dbpedia features possess co-linearity, LR approach will experience difficulty in modelling the relationship between the input and target variables.
\paragraph{Least Absolute Shrinkage and Selection Operator (LASSO)}
Co-variance defines the inter-dependence between the features in the dataset. A high-covariance feature fails to provide a precise prediction, whereas low-covariance features work better by adding diversification in the model. This results in more pattern learning and better returns. LASSO is one such type of linear regression model that aims at optimizing by keeping those features that provide low covariance. Due to this characteristic feature of LASSO, the model tends to overfit. In order to prevent the model from overfitting, we add a regularization parameter[15] obtained from AIC evaluation of LASSO.  We experimented with varying regularization values: \{2.0, 1.0, 0.50, 0.25,0.125 \}
\paragraph{ElasticNet}
In the current state of Twitter assembly, not every Twitter user has an extremely large follower count. Backpropagating to non-popular Twitter user from a popular Twitter user profile, generates sparse feature vectors. Such data fails to support variable selection in LASSO and Linear Regression, thus suffering from poor curve fitting. With this assumption, we realize the age prediction problem using ElasticNet. This linear model handles sparsity in the data with l1 penalization [28] and grouped variable selection (calculates collinearity between features). In our approach, we used ElasticNet in its vanilla states and compared it with the other linear models.
\paragraph{Support Vector Regression}
In the field of machine learning, Support Vector Machines (SVM) have been important in various classification tasks, such as drug-abuse population categorization, depression symptom classification, text classification, etc. Appreciating its robust performance in classification due to its ability to be modeled as linear (Linear Kernel) to non-linear (Radial Basis Function (RBF) Kernel), we intend to apply its regression type modeling to age prediction. In relation to linear function based regression, we tend to state Linear Kernel Support Vector Regressor (SVR) in this section. SVR identifies non-linearity in the data (low-variance, diversification) and provides a prediction model using features that constitute the non-linearity set. By default, SVR places an $epsilon$ threshold over residual error. As long as the error is less than $\epsilon$, it is neglected and model learning is not penalized.  

\subsection{Non-Linear Models}
Real world generation of data makes identification of distributional semantics of its feature values impossible[10]. For the realization of representation space of the model, we compared the efficacy of regression between SVR with a linear kernel versus a Radial Basis Function (RBF) kernel. These are Gaussian functions, that transform Cartesian coordinate to Gaussian space, so as to visualize non-linearity.  The polynomial kernel is another such kernel in SVR that transforms to the quadratic plane. We were interested in using this model because of the presence of non-linearity in data and high dimensional feature set. Hence, a radial basis kernel tries to map non-linear feature to high-dimensional space with an intuition that the features show linearity in the Gaussian space.  

\subsection{Interactional Feature Scaling}
% need to ask alan, what does he mean by interactional 
Features of the dataset are the characteristic fields that represent crowd (only if it includes users as its samples) behavior. An important point of consideration in machine learning is feature scaling which entails brings all the feature values to the same range. It is counter-intuitive when we neglect elementary mathematical relationship between the attributes and delegate the job to the model. Attributes in the dataset that requires modification by elementary arithmetic involving other feature(s) are termed as interactional features. In our dataset, DBpedia features are of this kind, seeing that we represent them as the count of a number of popular users who has a DBpedia mention. We investigate scaling of interactional features using \textit{popular follower count} and \textit{total follower count}. We performed normalization on remaining features.

\subsection{Normalization and Regularization}
Before, detailing more about the regularization and $\lambda's$, we define regularization.
\paragraph{Regularization} is a technique used to prevent the model from over-fitting by adding a high weighted term to the objective function. There are various ways of formulating these terms, such as  , L1 ($||\theta||$), L2 ($||\theta||^2_{2}$). We have used Frobenius norm [29] to formulate our optimization function with regularization. 

\begin{equation}\label{eq:1}
J(\theta) = \sum_{i=1}^{N_{tr}} (h(\theta)_{i} - Y_{i})^2  + \frac{\lambda}{2}||\theta||^2_{2}
\end{equation}
where 
$J(\theta)$ is the cost function, $\theta$ are the weights, \\
$N_{tr}$ : Number of the training samples. This value changes with the cross folds ratios. \\
$\lambda$: this is the regularization parameter. \\
$h(\theta)_{i}$ : this is the hypothesis function.\\ 
$Y_{i}$ : is the $i^{th}$ observation of \textit{age}.
Regularization and feature scaling are the techniques used to make the model reach its goal of convergence and better prediction without over-fitting and preventing the ill-posed problem.
In this experiment we varied values of $\lambda$ between 0.1 and 0.0001. This variation affected the accuracy rate of the model. 
\paragraph{Normalization}: Normalization is also called as Feature Scaling. We applied Min-Max normalization for our experiment. There are various ways for performing normalization such as Z-transformation, Scaling to unit length, Re-scaling,  and Standardization.
\begin{equation}\label{eq:2}
scaled-dataset =\frac{feature^{i}_{j} - \mu_{feature^i}}{max(feature^i)-min(feature^i)}
\end{equation}
i $\epsilon$ \{ selected features from our Twitter dataset\} and j $\epsilon$ [0, length($feature^{i}$)-1]

\subsection{Evaluation Metrics}
 We assess the efficacy of the models using four different metrics: mean absolute error, median absolute error, $R^2$-score and Accuracy@10. Each metric is representative of model performance and is also influenced by the nature of the data. Mean Absolute Error (MAE) is defined as the averaged deviations of prediction value from the true value. MAE is sensitive to outliers (in the target variable) as they generate high deviation. In contrast, Median Absolute Error (MedAE) is indifferent towards outliers, as it takes the median of the deviations. Assessing the model using both MAE and MedAE is important to define the model performance in the presence/absence of skewed data. Moreover, the values of these error metrics provide an approximation of the standard deviation in the prediction error. These model evaluation measures have seen utility in classification tasks where the prediction is more of an estimation of age-classes as opposed to precise judgment. In order to gauge the model precision, we harness $R^2$-score and Accuracy@10 metrics. $R^2$-score is also called the coefficient of determination[22]. It is a measure of closeness to the regression line. A myth with respect to this metric is that high $R^2$-score is always better over low score[23]. However, its importance is appreciated along with MAE, MedAE and Accuracy@10. Another metric for developing a sense of model's precision is Accuracy@10. We define Accuracy@10, as the percentage accuracy of predicted age given an error window of $\pm$10. For instance, a user of age 20, its window of 10 is [10,30]. A prediction of 24 by the model is considered as a correct prediction by this metric. In order to quantify the model's precise judgment, we state Accuracy@\{2,5,7\} (only for the model with the best performance at Accuracy@10).

\section{Results and Analysis}
% Write about Residual plots
% Write Sigmoid plot
% list top 5 best model performance with MAE, Median Error and R2-score, comparing with the base line
% (Not sure) we provide the best model prediction Accuracy@2, 5, 7
In this section, we discuss a series of experiments that were performed to evaluate the efficacy and preciseness of the model based on presence and absence of interactional feature scaling.

\subsection{Analysis without Interaction Feature Scaling}
Feature scaling is a part of the feature engineering process that unifies the range across all the features in the dataset. But, the requirement of feature scaling over all features has been debated. When we do not perform interaction feature scaling and give the model a preprocessed data (removed outliers, missing values etc.). For this we investigated 5 linear and 1 non-linear regressor (SVR-RBF). According to the table \ref{table:1}, we observed that SVR-RBF performed well in comparison to baseline and SVR-L and LASSO model regularized at $\lambda = 0.25$. An improvement over MAE and MedAE for the SVR-RBF and other comparable models does not define precision of the model. An extension to the evaluation we employed $R^2$-score and Accuracy@10 measure. We noted two interesting findings; First, $R^2$-score was less for SVR-RBF. Second, Accuracy@10 for SVR-RBF was better over all the models. We put forward SVR-RBF as the best model overall because of 2 reasons; first, Mean and Median Absolute Errors were lowest. Second, SVR-RBF provided more precise prediction within the range [-10,+10] of the true age. 

\begin{table}[!htbp]
\centering
\begin{tabular}{|p{3cm}|p{1.2cm}|p{1.2cm}|p{1.2cm}|p{1.2cm}|  }
\hline
\multirow{2}{*}{Model} &\multicolumn{4}{|c|}{Evaluation}\\ \cline{2-5}
& MAE &MedAE & $R^2$ & Acc.@10\\
\hline
SVR-RBF & 6.8 & 3.2 & 0.02 & 0.84\\
\hline
SVR-L & 6.9 & 3.3 & 0.004 & 0.83\\
\hline
LASSO@0.25 & 7.6 & 4.9 & 0.08 & 0.81 \\
\hline
Baseline (mean) & 8.1 & 5.8 & 0.0 & 0.81 \\
\hline
\end{tabular}
\caption{Age Prediction performance measurement using MAE, MedAE, $R^2$-score and Accuracy@10 in the absence of interactional feature scaling. SVR-RBF: Support Vector Regression using Radial Basis Function Kernel, SVR-L: Support Vector Regression using Linear Kernel, Baseline: Predicting mean value. Test Set of 7630 Users, Trainset: 15490. This is same for table\ref{table:2}} 
\label{table:1}
\end{table}

\subsection{Analysis with Interaction Feature Scaling}
Though we performed feature scaling in all our experiments, in this section we present the results of using an interaction feature between each of the DBpedia features and the user's popular friends counts. We replace the raw value (count) of each DBpedia feature with that value divided by the number of popular friends the user follows. The intuition behind this is that the relative proportions of each type of friend a user follows should yield a higher correlation with their age than would the raw counts of each. So, we "scale" these features on a per-observation basis, thereby normalizing each to be a percentage of the user's total popular friends.

\begin{figure}[!htbp]
\centering
\includegraphics[scale=0.50]{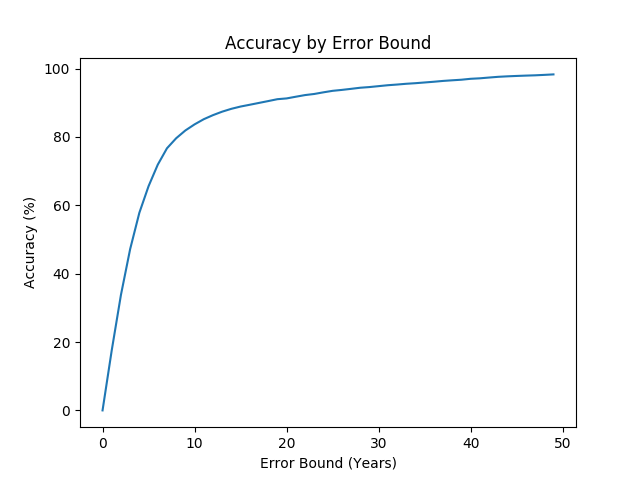}
\caption{Fine-grained error bound to coarse grained error bound. An age-range based precision analysis of Support Vector Regressor with Radial Basis Kernel}
\label{fig:3}
\end{figure}

\begin{figure}[!htbp]
\centering
\includegraphics[scale=0.50]{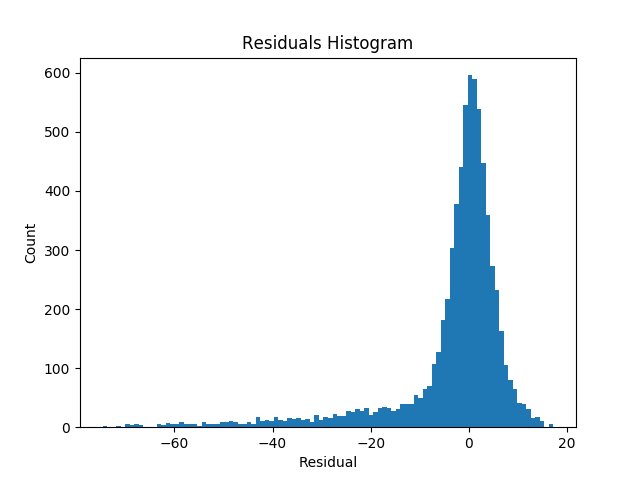}
\caption{Near left skewed Residual plot of 7630 test users. Majority of error is centered around zero. }
\label{fig:4}
\end{figure}

\begin{table}[!htbp]
\centering
\begin{tabular}{ |p{3cm}|p{1.2cm}|p{1.2cm}|p{1.2cm}|p{1.2cm}|  }
\hline
\multirow{2}{*}{Model} &\multicolumn{4}{|c|}{Evaluation}\\ \cline{2-5}
 & MAE &MedAE & $R^2$ & Acc.@10\\
\hline
SVR-RBF & 6.8 & 3.2 & 0.03 & 0.84\\
\hline
LASSO@1 & 7.9 & 5.2 & 0.05 & 0.80\\
\hline
LASSO@0.5 & 7.7 & 4.9 & 0.09 & 0.80 \\
\hline
LASSO@0.25 & 7.6 & 4.8 & 0.10 & 0.79 \\
\hline
LASSO@0.125 & 7.5 & 4.7 & 0.11 & 0.79 \\
\hline
Baseline (mean) & 8.1 & 5.8 & 0.0 & 0.81 \\
\hline
\end{tabular}
\label{table:2}
\caption{Age Prediction performance measurement using MAE, MedAE, $R^2$-score and Accuracy@10 in the presence of interactional feature scaling. SVR-RBF: Support Vector Regression using Radial Basis Function Kernel, LASSO@$\lambda$ model with regularization; $\lambda$ $\epsilon$ \{1,0.5,0.25,0.125\} } 
\end{table}
Though our models provided low value for coefficient of determination ($R^2$-score, intuitive due to noise in the Twitter dataset) we are able to show that the model is able to predict exact age of 2\% of the new users and provide 83\% approximate prediction of age for new users. Now, we assess the efficacy of our best model - SVM-RBF over the entire population. Figure \ref{fig:3}, provide the preciseness in the prediction of age when we move from fine grained to coarse grained window of error in age. For instance; an age range of $\pm4$ is fine grained and an age range of $\pm20$ is coarse grained. We observed that the model does better as the width the age range (error bound) increases. Keeping a threshold of 10, we observed that model is able to provide a correct prediction for 83\% of the users (in the test set) which is 6333 users out of 7630. 
In the figure\ref{fig:3}, there is flattening of the curve when we move from fine error bound to coarse error bound, which we visualized using the residual plots. From the figure \ref{fig:4}, we observe that our model is predicting the age of a user less than his/her true age majority of the time. Resulting are residual plot to be left skewed. However, high density around zero marks the fairly good performance of the model on the dataset with statistical and knowledge-based features. The plot of predicted versus expected user age testify the density around zero in the residual plot. Presence of skewness and sparsity has driven the model more close to the baseline prediction but still providing better performance on the Accuracy@10 measure (see table \ref{table:4}). As a part of future work, we plan to collapse some DBpedia categories to more abstract categories and subsequently evaluate the relevance of skewed age population. Moreover, we plan to remove some of the DBpedia categories that are rarely followed by non-popular twitter users (i.e. sparse features). Based on our series of experiments, we define our future road map in subsequent section.  
\begin{figure}[!htbp]
\centering
\includegraphics[scale=0.50]{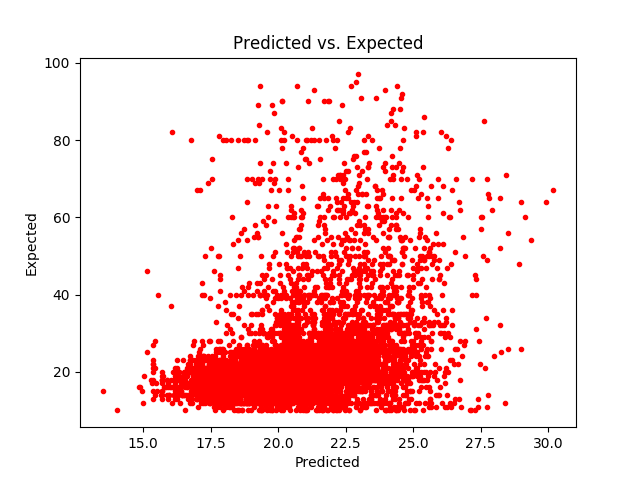}
\caption{Predicted versus Expected age plot to analyze the goodness of prediction of SVR-RBF for younger, adolescent and old age twitter users.}
\label{fig:5}
\end{figure}

\begin{table}[!htbp]
    \centering
    \begin{tabular}{|c|c|}
    \hline
    Error-Bound & Accuracy\\
    \hline
        0 &0.000000\\
        \hline
        1 &0.176046\\
        \hline
        2 &0.338094\\
        \hline
        3 &0.472002\\
        \hline
        4 &0.578196\\
        \hline
        5 &0.655967\\
        \hline
        6 &0.718891\\
        \hline
        7 &0.766827\\
        \hline
        8 &0.796239\\
        \hline
        9 &0.819005\\
        \hline
        10 &0.836821\\
    \hline
    \end{tabular}
    \caption{Accuracy@Error-Bound results of SVR-RBF. Error-Bound is the age-range. For instance, Error-bound of 2 means that regressor can predict the age with an error of $\pm$2. Accuracy@Error-Bound turns to Accuracy@10 in our explanation.}
    \label{table:4}
\end{table}

\section{Conclusion and Future Work}
Based on the investigation with and without interactional feature pre-scaling, it is conclusive that SVR-RBF remain indifferent whereas LASSO  and linear kernel SVR showed a high rate of variations in Accuracy@10. Moreover, we observed a consistency in performance (MAE, Accuracy@10) over younger, adolescent and older age groups. Moreover, our low $R^2$-score value showes high sparsity in the data. Hence we seek to collapse some of the DBpedia categories which are not representative of the large population. 
We intend to perform a task of classification by segregating the user at the median age of the data. We seek to perform regression for age prediction over users having age greater than median and grouped by age (e.g if median is 25, then proposed groups are 25+, 35+ and 45+). Furthermore, we propose the use of content based features together with statistical and DBpedia features to meliorate the age prediction task. We foresee using rank correlation over the dataset as a way to impose more strict feature selection to overcome sparsity problem.  

%\section{Distribution of Work}
%During the course of our machine learning project, we came across many challenges, limitations and knowledge bottlenecks. Overcoming them has never been a trivial task. As a team focused towards prototyping an innovation in the domain of social media analysis,  we required a distributive yet collaborative approach. That being said, it is hard to demarcate the distribution of the work as each of us has extremely contributed towards the success of our prototyping. Alan has been involved in data collection stage, making crawlers for Twitter and DBpedia knowledge graph. Manas showed inclination towards initial data analysis, and the initial selection and justification of candidate algorithms. We worked together closely to define the appropriate evaluation metrics, and to interpret the changes in each with respect to our problem domain. We collaborated in-depth on our residual analysis, and overall brainstorming and tweaking of the dataset in our attempts to improve the model. We worked closely to design the structure of the iterative execution of the machine learning pipeline and hyper-parameter tuning grid.  

\section{References}
\begin{enumerate}
    \item Westenberg, W. M. The influence of Youtubers on teenagers: a descriptive research about the role Youtubers plays in the life of their teenage viewers. MS thesis. University of Twente, 2016.
    \item Hu, Jian, et al. "Demographic prediction based on user's browsing behavior." Proceedings of the 16th international conference on World Wide Web. ACM, 2007.
    \item Guo, Guodong, et al. "A probabilistic fusion approach to human age prediction." Computer Vision and Pattern Recognition Workshops, 2008. CVPRW'08. IEEE Computer Society Conference on. IEEE, 2008.
    \item Nguyen, Dong, et al. "Why gender and age prediction from tweets is hard: Lessons from a crowdsourcing experiment." Proceedings of COLING 2014, the 25th International Conference on Computational Linguistics: Technical Papers. 2014.
    \item Delip Rao, David Yarowsky, Abhishek Shreevats, and Manaswi Gupta. 2010. Classifying latent user attributes in twitter. In Proceedings of the 2nd international workshop on Search and mining user-generated contents (SMUC '10). ACM, New York, NY, USA, 37-44. 
    \item Nguyen, Dong-Phuong, et al. ""How old do you think I am?" A study of language and age in Twitter." (2013).
    \item Zhuang, Dong, et al. "Demographic prediction using a social link network." U.S. Patent Application No. 11/535,160.
    \item Nguyen, Dong, Noah A. Smith, and Carolyn P. Rosé. "Author age prediction from text using linear regression." Proceedings of the 5th ACL-HLT Workshop on Language Technology for Cultural Heritage, Social Sciences, and Humanities. Association for Computational Linguistics, 2011.
    \item Tibshirani, Robert. "Regression shrinkage and selection via the lasso." Journal of the Royal Statistical Society. Series B (Methodological) (1996): 267-288.
    \item Matwin, Stan, and Jan Mielniczuk, eds. Challenges in Computational Statistics and Data Mining. Springer International Publishing, 2016.
    \item Massey Jr, Frank J. "The Kolmogorov-Smirnov test for goodness of fit." Journal of the American Statistical Association 46.253 (1951): 68-78.
    \item Ryder, Norman B. "The cohort as a concept in the study of social change." American sociological review (1965): 843-861
    \item “Women's Health.” World Health Organization, World Health Organization, Sept. 2013, www.who.int/mediacentre/factsheets/fs334/en/.
    \item Jadhav, Ashutosh Sopan, et al. "Twitris 2.0: Semantically empowered system for understanding perceptions from social data." (2010).
    \item Friedman, Jerome H. "Regularized discriminant analysis." Journal of the American statistical association 84.405 (1989): 165-175.
    \item Terry, Mark. "Twittering healthcare: social media and medicine." Telemedicine and e-Health 15.6 (2009): 507-510.
    \item Hambrick, Marion E., et al. "Understanding professional athletes’ use of Twitter: A content analysis of athlete tweets." International Journal of Sport Communication 3.4 (2010): 454-471.
    \item Vieweg, Sarah, et al. "Microblogging during two natural hazards events: what twitter may contribute to situational awareness." Proceedings of the SIGCHI conference on human factors in computing systems. ACM, 2010.
    \item Goel, Sharad, et al. "Predicting consumer behavior with Web search." Proceedings of the National academy of sciences 107.41 (2010): 17486-17490.
    \item Kahne, Joseph, Nam-Jin Lee, and Jessica T. Feezell. "The civic and political significance of online participatory cultures among youth transitioning to adulthood." Journal of Information Technology \& Politics 10.1 (2013): 1-20.
    \item Buhrmester, Michael, Tracy Kwang, and Samuel D. Gosling. "Amazon's Mechanical Turk: A new source of inexpensive, yet high-quality, data?." Perspectives on psychological science 6.1 (2011): 3-5.
    \item Ozer, Daniel J. "Correlation and the coefficient of determination." Psychological Bulletin 97.2 (1985): 307.
    \item Frost, Jim. “Regression Analysis: How Do I Interpret R-Squared and Assess the Goodness-of-Fit?” Minitab, 30 May 1970, blog.minitab.com/blog/adventures-in-statistics-2/regression-analysis-how-do-i-interpret-r-squared-and-assess-the-goodness-of-fit.
    \item Amit Sheth, Hemant Purohit, Gary Alan Smith, Jeremy Brunn, Ashutosh Jadhav, Pavan Kapanipathi, Chen Lu, Wenbo Wang. Twitris: A System for Collective Social Intelligence. In: Reda Alhajj, Jon Rokne. Encyclopedia of Social Network Analysis and Mining. 2nd ed. New York: Springer-Verlag New York; 2018. p. 1-23.
    \item Twitter, 2017. API Reference Index, \{https://developer.twitter.com/en/docs/api-reference-index\} (accessed on Web Page 2017).
    \item DBpedia, 2017. DBpedia Version 2016-10. \{http://wiki.dbpedia.org/downloads-2016-10\#datasets\} (accessed on Web Page 2017).
    \item Wikipedia, 2017. List of most-followed Twitter accounts,{https://en.wikipedia.org/wiki/List\_of\_most-followed\_Twitter\_accounts} (accessed on Web Page 2017).
    \item Culotta, Aron, Nirmal Ravi Kumar, and Jennifer Cutler. "Predicting Twitter User Demographics using Distant Supervision from Website Traffic Data." J. Artif. Intell. Res.(JAIR) 55 (2016): 389-408.
    \item Mnih, Andriy, and Ruslan R. Salakhutdinov. "Probabilistic matrix factorization." Advances in neural information processing systems. 2008.
    
\end{enumerate}
\end{document}